\documentclass[aps,pra,twocolumn,preprintnumbers,amsmath,amssymb,
superscriptaddress,floatfix,a4paper]{revtex4-2}

\usepackage{graphicx}

 \usepackage{amsmath}
 \usepackage{amssymb}
\usepackage[dvipsnames]{xcolor}
\usepackage{physics}

\usepackage[utf8]{inputenc}
\usepackage{siunitx}

\usepackage[hidelinks]{hyperref}

\hypersetup{
colorlinks=true,    
linkcolor=blue,
citecolor=blue,
urlcolor=blue,
pdfborder={0 0 0},
bookmarksopen=true,
bookmarksopenlevel=2,
pdfpagemode=UseOutlines,
pdfstartview={Fit},
pdftitle={Controllable orbital angular momentum monopoles in chiral topological semimetals},
pdfauthor={J. A. Krieger, Y. Yun, M. Yao},
pdfsubject={OAM chiral topological semimetals},
pdfkeywords={PtGa,PdGa, CD-ARPES, simulated photoemission},
pdfhighlight= /I
}

\newcommand{\refsubfig}[2]{\hyperref[#1]{\ref*{#1}#2}}

\renewcommand{\vec}[1]{\mathbf{#1}}

\begin{document}

\let\oldaddcontentsline\addcontentsline
\renewcommand{\addcontentsline}[3]{}

\title{Controllable orbital angular momentum monopoles in chiral topological semimetals}

\author{Yun~Yen}
\altaffiliation{These authors contributed equally to this work}
\affiliation{Laboratory for Materials Simulations, Paul Scherrer Institute, Villigen PSI, Switzerland}
\affiliation{École Polytechnique Fédérale de Lausanne (EPFL), Switzerland}

\author{Jonas~A.~Krieger}
\altaffiliation{These authors contributed equally to this work}
\affiliation{Max Planck
Institut f\"ur
Mikrostrukturphysik,
Weinberg 2, 06120 Halle, Germany}
\author{}
\altaffiliation[Current address: ]{Laboratory for Muon Spin Spectroscopy, Paul Scherrer Institute, CH-5232 Villigen PSI, Switzerland}
\noaffiliation
\author{Mengyu~Yao}
\altaffiliation{These authors contributed equally to this work}
\affiliation{Max Planck Institute for Chemical Physics of Solids, Dresden, Germany}

\author{Iñigo~Robredo}
\affiliation{Max Planck Institute for Chemical Physics of Solids,
Dresden, Germany}
\affiliation{Donostia International Physics Center,
20018 Donostia - San Sebastian, Spain}

\author{Kaustuv~Manna}
\affiliation{Max Planck Institute for Chemical Physics of Solids,
Dresden, Germany
}
\affiliation{Indian Institute of Technology-Delhi, Hauz Khas, New Delhi 110 016, India}

\author{Qun~Yang}
\affiliation{Max Planck Institute for Chemical Physics of Solids,
Dresden, Germany
}

\author{Emily~C.~McFarlane}
\affiliation{Max Planck
Institut f\"ur
Mikrostrukturphysik,
Weinberg 2, 06120 Halle, Germany}

\author{Chandra~Shekhar}
\affiliation{Max Planck Institute for Chemical Physics of Solids,
Dresden, Germany
}

\author{Horst~Borrmann}
\affiliation{Max Planck Institute for Chemical Physics of Solids,
Dresden, Germany
}

\author{Samuel~Stolz}
\affiliation{nanotech@surfaces Laboratory, Empa,
Swiss Federal Laboratories for Materials Science and Technology,
8600 Dübendorf, Switzerland}

\author{Roland~Widmer}
\affiliation{nanotech@surfaces Laboratory, Empa,
Swiss Federal Laboratories for Materials Science and Technology,
8600 Dübendorf, Switzerland}

\author{Oliver~Gröning}
\affiliation{nanotech@surfaces Laboratory, Empa,
Swiss Federal Laboratories for Materials Science and Technology,
8600 Dübendorf, Switzerland}

\author{Vladimir~N.~Strocov}
\affiliation{Photon Science Division,
Paul Scherrer Institute,
5232 Villigen PSI, Switzerland}

\author{Stuart~S.P.~Parkin}
\affiliation{Max Planck Institut f\"ur
Mikrostrukturphysik, Weinberg 2, 06120 Halle, Germany}

\author{Claudia~Felser}
\email{claudia.felser@cpfs.mpg.de}
\affiliation{Max Planck Institute for Chemical Physics of Solids, Dresden, Germany}

\author{Maia~G.~Vergniory}
\affiliation{Max Planck Institute for Chemical Physics of Solids, Dresden, Germany}

\affiliation{Donostia International Physics Center,
20018 Donostia - San Sebastian, Spain}

\author{Michael~Sch\"uler}
\email{michael.schueler@psi.ch}
\affiliation{Laboratory for Materials Simulations, Paul Scherrer Institute, Villigen PSI, Switzerland}
\affiliation{Department of Physics, University of Fribourg, Fribourg, Switzerland}

\author{Niels~B.~M.~Schr\"oter}
\email{niels.schroeter@mpi-halle.mpg.de}
\affiliation{Max Planck Institut f\"ur
Mikrostrukturphysik, Weinberg 2, 06120 Halle, Germany}

\begin{abstract}
The emerging field of orbitronics aims at generating and controlling currents of electronic orbital angular momentum (OAM) for information processing. Structurally chiral topological crystals could be particularly suitable orbitronic materials because they have been predicted to host topological band degeneracies in reciprocal space that are monopoles of OAM. Around such a monopole, the OAM is locked isotopically parallel or antiparallel to the direction of the electron’s momentum, which could be used to generate large and controllable OAM currents. However, OAM monopoles have not yet been directly observed in chiral crystals, and no handle to control their polarity has been discovered. Here, we use circular dichroism in angle-resolved photoelectron spectroscopy (CD-ARPES) to image OAM monopoles in the chiral topological semimetals PtGa and PdGa. Moreover, we also demonstrate that the polarity of the monopole can be controlled via the structural handedness of the host crystal by imaging OAM monopoles and anti-monopoles in the two enantiomers of PdGa, respectively. For most photon energies used in our study, we observe a sign change in the CD-ARPES spectrum when comparing positive and negative momenta along the light direction near the topological degeneracy. This is consistent with the conventional view that CD-ARPES measures the projection of the OAM monopole along the photon momentum. For some photon energies, however, this sign change disappears, which can be understood from our numerical simulations as the interference of polar atomic OAM contributions, consistent with the presence of OAM monopoles. Our results highlight the potential of chiral crystals for orbitronic device applications, and our methodology could enable the discovery of even more complicated nodal OAM textures that could be exploited for orbitronics.

\end{abstract}

\maketitle
\noindent

\section*{Introduction}
The motion of electrons in solids can be described by wave packets that may possess a self-rotation around their centre of mass. This self-rotation gives rise to an orbital magnetic moment that is proportional to the orbital angular momentum (OAM) of the Bloch states~\cite{souza_dichroic_2008,xiao_berry_2010}. Currents of orbital angular momentum have been proposed as a viable alternative to spin-currents to manipulate and control magnetism in nanoscale memory devices by giving rise to large spin- and orbital-torques~\cite{bernevig_orbitronics_2005,go_orbitronics_2021,go_long-range_2023,hayashi_observation_2023,bose_detection_2023,rappoport_first_2023}. However, most recent investigations of orbital currents and the related orbital Hall effect have mainly focused on non-magnetic centrosymmetric materials where OAM is suppressed in the equilibrium ground state, and can only be generated under the application of an electrical field or at surfaces~\cite{park_orbital-angular-momentum_2011,go_toward_2017,choi_observation_2023}. Magnetic materials with their intrinsic orbital moment have also been shown to exhibit the orbital Hall effect~\cite{sala_orbital_2023,lyalin_magneto-optical_2023}.
The common key feature for a sizable orbital Hall effect is a
nontrival momentum dependence of magnetic orbitals (known as orbital texture~\cite{han_theory_2023}), which can enable orbital currents even in centrosymmetric crystals~\cite{go_intrinsic_2018,cysne_disentangling_2021}.
In nonmagnetic helical molecules and nonmagnetic structurally chiral crystals, the Bloch bands already possess such an OAM texture in the ground state, which will lead to an orbital polarization when an electric field is applied that could be exploited for launching and injecting large orbital currents in orbitronic devices.
In helical molecules, this orbital polarization has recently been attributed to the observation of the chiral-induced spin selectivity (CISS) effect~\cite{naaman_chiral_2020} due to the conversion of OAM to spin in contacts with strong spin-orbit coupling ~\cite{liu_chirality-driven_2021}. However, in contrast to chiral molecules, much less is known experimentally about the OAM textures in chiral crystals.

In chiral cubic crystals, crystalline symmetries can stabilize two-fold and multifold degenerate band crossings at time-reversal invariant momenta of the Brillouin zone, which are known as Kramers-Weyl- and multifold-fermions~\cite{bradlyn_beyond_2016,chang_topological_2018}.
These crossing points are expected to be strong sources and sinks of OAM, which results in large orbital Hall effects and current-induced orbital magnetization with a susceptibility that is predicted to be an order of magnitude larger than in corresponding Rashba systems~\cite{he_kramers_2021,yang_monopole-like_2023}. Moreover, due to the high symmetry of cubic crystals and the absence of mirror symmetries in chiral systems, the direction of the OAM is expected to be locked isotropically parallel or antiparallel along the direction of the crystal momentum of the Bloch states in the vicinity of a linear band crossing point. As a result,
the induced OAM currents and orbital magnetization will only depend on the direction of the injected charge current, and not on the crystallographic direction along which the charge current is flowing. Such an isotropic longitudinal magnetoelectric effect could be used for novel device applications, for instance for the switching of magnetic domains with perpendicular magnetic anisotropy~\cite{he_kramers_2021}. Since rotational disorder of crystal grains is common in thin-film devices that could quench an anisotropic response, OAM monopoles with isotropic parallel OAM locking could be particularly valuable for device applications. Since OAM is a pseudovector, an OAM monopole will be transformed into an anti-monopole and vice versa upon reversing the structural chirality of the host crystal. Demonstrating this transformation law could highlight structural chirality as a design parameter for orbitronic applications because the polarity of the OAM monopole determines the directionality of transport responses. More generally, OAM monopoles are a manifestation of the quantum geometry of the electrons in periodic solids, which influences Fermi-liquid transport~\cite{karplus_hall_1954,neupert_measuring_2013} and opto-electronic properties~\cite{Yao08, ahn_riemannian_2022,topp_light-matter_2021-1,morimoto_topological_2016-1,puente-uriona_ab_2023}.

Multifold fermions have recently been observed with angle-resolved photoelectron spectroscopy (ARPES) in cubic chiral crystals of the B20 crystal structure ~\cite{takane_observation_2019,sanchez_topological_2019,schroter_chiral_2019,schroter_observation_2020}. However, these experiments only investigated the dispersion relationship of these multifold fermions, but not their OAM texture. Therefore, so far, direct experimental evidence for the existence of OAM monopoles near Kramers-Weyl or multifold fermions in chiral crystals remains elusive. Moreover, it has not yet been demonstrated that the polarity of such OAM monpoles can be controlled via the structural chirality of the host crystal. Note that achiral Weyl-semimetals such as TaAs can also host OAM monopoles ~\cite{unzelmann_momentum-space_2021}. However, due to the mirror symmetries present in these compounds, monopoles with opposite polarities are pinned to the same energies, which can lead to a cancellation of the effective orbital polarization, which makes them less valuable for orbitronics. In contrast, in chiral topological semimetals, monopoles with opposite polarity are generically separated in energy ~\cite{schroter_chiral_2019}, which has been predicted to result in large orbital polarizations~\cite{he_kramers_2021,yang_monopole-like_2023}.  Moreover, because band degeneracies cannot be pinned to high symmetry points in achiral Weyl semimetals, they cannot exhibit isotropic OAM momentum locking as in the case of chiral topological semimetals.

\begin{figure*}[t]
    \includegraphics[width=1.0\textwidth]{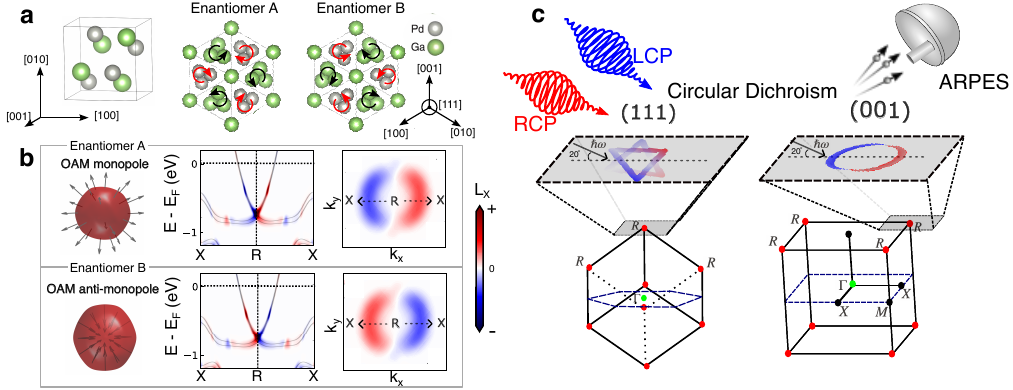}
    \caption{\textbf{Orbital angular momentum monopoles in chiral topological semimetals PdGa / PtGa.} \textbf{a} Sketch of the chiral crystal structure of PdGa (B20 structure), highlighting the helical winding of the atoms along the [111] crystal direction. \textbf{b} Calculated OAM monopole texture around the double spin-1 multifold band degeneracy at the R-point (left) in the corner of the Brillouin zone. The arrows on the sphere indicate the OAM direction on the constant energy contour at 30 meV above the crossing.  Parallel OAM-momentum locking along the X-R-X direction (middle) and on the isoenergy surface in the X-R-M plane $\sim$ 0.1 eV above the node (right).  \textbf{d} Illustration of the experimental geometry of the CD-ARPES experiment. Varying the crystal orientation allows us to probe the three-dimensional OAM texture along different directions around the R-point.}
    \label{fig:intro}
\end{figure*}

Probing the OAM of Bloch electrons -- with its full momentum dependence to identify the presence of OAM monopoles -- still remains a grand challenge. Since spin and orbital moments in solids influence the circular dichroism observed in many spectroscopic probes, such as X-ray circular dichroism (XMCD), it seems natural to suspect that circular dichroism in ARPES (CD-ARPES) might be used to map out the momentum distribution of the OAM of Bloch states. Indeed, for surface states of bulk Au~\cite{park_orbital-angular-momentum_2011} and Bi$_2$Se$_3$~\cite{Wang11,park_chiral_2012} as well as for two-dimensional materials~\cite{Cho18,Schuler20}, a qualitative agreement between CD-ARPES and the OAM (projected onto the incidence direction of light) has been observed.
While this apparent direct link from circular dichroism to the OAM texture, which is in turn closely related to the Berry curvature~\cite{xiao_berry_2010}, is the working hypothesis of recent experiments~\cite{cho_studying_2021,unzelmann_momentum-space_2021,di_sante_flat_2023}, its general validity for bulk materials is far from clear. Details of the experimental geometry and complicated final states~\cite{strocov_high-energy_2023,kern_simple_2023} obscure the link. Even for simple materials where the bands are dominated by $d$ orbital character, the circular dichroism is not necessarily indicative of the OAM~\cite{Schonhense90,moser_toy_2023}. For materials with multiple atoms per unit cell, (photon-energy-dependent) interference effects further complicate the signal~\cite{jiang_observation_2013,heider_geometry-induced_2023}. Hence, to understand complex quantum materials -- such as chiral topological semimetals hosting OAM monopoles -- a microscopic picture beyond the standard paradigm is required.

In this paper, we fill this gap for the chiral topological metals PdGa and PtGa with a blend of CD-ARPES experiments and photoemission simulations. Our calculations are based on a unique compromise between accurately reproducing the measured signal and transparency of our model, which enables us to interpret the complicated photon-energy dependence of our CD-ARPES spectra, revealing the first direct evidence for the presence of OAM monopoles in these chiral crystals.

\section*{Results}

\begin{figure*}
    \includegraphics[width=1.0\textwidth]{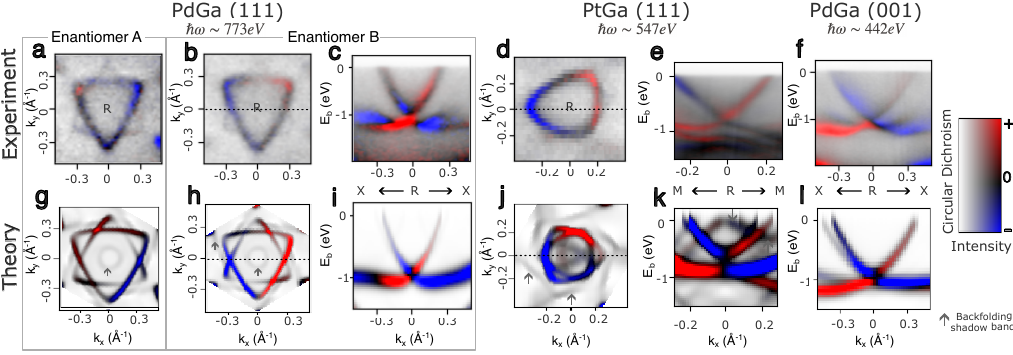}
    \caption{ \textbf{Circular dichroism of angle-resolved photoemission near the R point in PdGa and PtGa.}
    The projection of the photon momentum on the sample surface is aligned with the $k_x$ direction. \textbf{a} and \textbf{b} shows dichroic Fermi surface data for PdGa(111) of two enantiomers, and \textbf{c} shows the binding energy dependent dichroic spectrum along the k dashed line in \textbf{b}. \textbf{d} and \textbf{e} shows the data for PtGa(111), and \textbf{f} shows the data for PdGa(001). Note that the $k_x$ direction in \textbf{e}(\textbf{b,f}) is aligned along MRM (XRX). \textbf{g}--\textbf{l} shows the corresponding calculated CD-ARPES spectra under the same conditions as in the experiments. Note that the two-dimensional colormaps encodes both the photoemission intensity and circular dichroism. The [111]- supercell calculation includes "shadow bands", artifacts due to the finite size of the slab geometry. These artifacts are indicated with gray arrows.}
    \label{fig:cd_orientation}
\end{figure*}

PdGa and PtGa crystallize in the structurally chiral B20 crystal structure (Space group 198, Fig.~\ref{fig:intro}\textbf{a}). The combination of crystalline symmetries and time-reversal symmetry protect a double spin-1 multifold fermion band crossing that is pinned to the R point in the corner of the cubic Brillouin zone~\cite{bradlyn_beyond_2016,chang_unconventional_2017,tang_multiple_2017}. For momenta close to the node, the Fermi surface is approximately spherical with an isotropic parallel locking between OAM and crystal momentum, which can be observed in ab-initio calculations of the OAM (see Fig.~\ref{fig:intro}\textbf{b-c}). As two enantiomers are structurally related to each other with mirror symmetry, the corresponding OAM monopoles show opposite polar texture.

\subsection*{Signarue of orbital texture in circular dichroism of angle-resolved photoemission}
To probe this polar OAM texture in PdGa and its related sister compound PtGa, we measured CD-ARPES Fermi-surfaces on the (001) and (111) crystal planes (corresponding measurement geometries sketched in Fig.~\ref{fig:intro}\textbf{d}). Following the traditional argument that CD-ARPES measures a projection of the OAM along the direction of the incident light, one would expect that the Fermi-surface maps near the R point should display a polar CD-ARPES texture, i.e. display a sign change for positive and negative momenta with respect to the R point, irrespective of the orientation of the crystal structure relative to the light direction or the employed photon energy.

In Fig.~\ref{fig:cd_orientation} we present the CD-ARPES measurements for the (111) surface of PtGa and PdGa (Fig.~\ref{fig:cd_orientation}\textbf{a}--\textbf{e}) and for the (001) surface of PdGa (Fig.~\ref{fig:cd_orientation}\textbf{f}). The incoming light momentum has a 20-degree grazing incidence angle with the $k_x$ axis and is orthogonal to the $k_y$ axis.  All measured spectra are accompanied by simulated CD-ARPES spectra (Fig.~\ref{fig:cd_orientation}\textbf{g}--\textbf{l}) based on the Wannier approach to ARPES~\cite{day_computational_2019,beaulieu_unveiling_2021,schuler_polarization-modulated_2022} (see Methods). While the complications of the final states are captured only approximately, the Wannier-ARPES method has been shown to qualitatively reproduce linear and circular dichroism for a range of systems.  The photon energies were carefully chosen such that the probed out-of-plane momentum corresponds to the plane containing the R-point in the three-dimensional cubic Brillouin zone.

As Fig.~\ref{fig:cd_orientation} demonstrates, a polar texture of circular dichroism near the multifold fermion band crossing at the R point is ubiquitous throughout the different materials and orientations. In other words, there is a characteristic sign change of the CD-ARPES spectrum for positive and negative $k_x$ momenta with respect to the R point. This observation is consistent with the conventional view that CD-ARPES measures the projection of the OAM monopole along the photon momentum. Moreover, this sign change is reversed when comparing the Fermi-surfaces of the two enantiomers in Fig.~\ref{fig:cd_orientation}\textbf{a}--\textbf{b}, which is consistent with the expectation that the polarity of the OAM pole can be controlled via the handedness of the host crystal (see supplemental material for more details~\cite{supplement}).

\subsection*{Photon-energy dependence of the circular dichroism}
\begin{figure*}
    \includegraphics[width=1.0\textwidth]{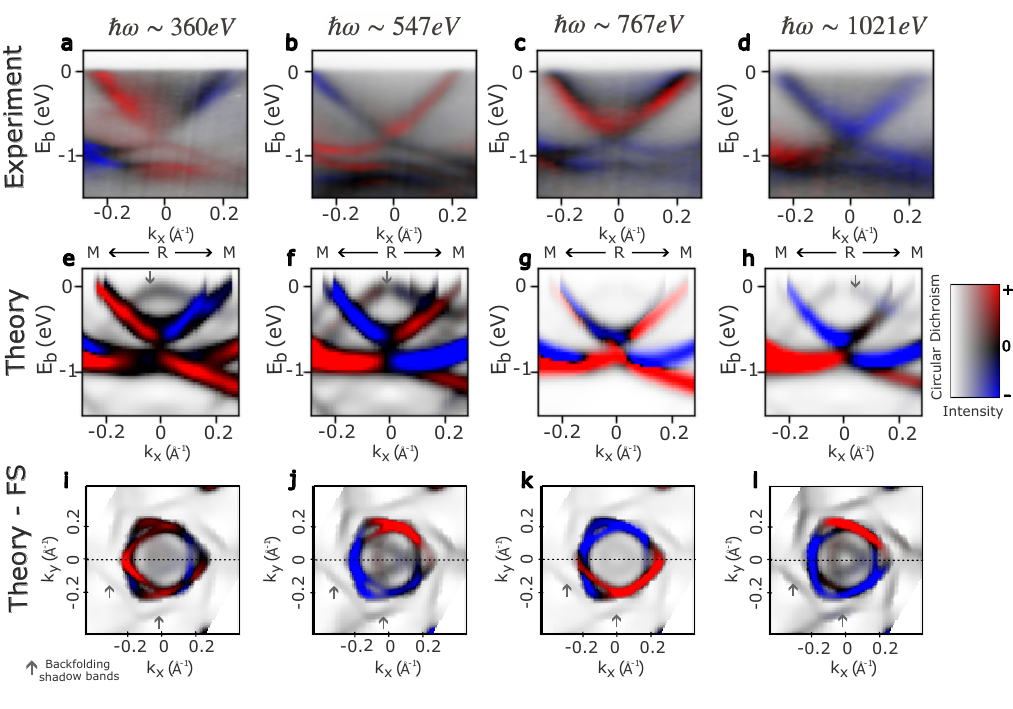}
    \caption{ \textbf{Photon-energy evolution of CD-ARPES spectra for PtGa (111).}
    \textbf{a}-\textbf{d} Energy vs. momentum resolved CD-ARPES along MRM directions acquired with different photon energies. \textbf{e}-\textbf{h} The corresponding calculated spectra and \textbf{i}-\textbf{l} the calculated CD-ARPES Fermi surfaces. The dashed lines in \textbf{i}-\textbf{l} correspond to the line cuts in \textbf{e}-\textbf{h}. The  backfolded "shadow band" artifacts have visible intensity due to the finite slab size.}
    \label{fig:cd_photon}
\end{figure*}

The simulated CD-ARPES spectra, which are in excellent qualitative agreement with the experiments, further support the robustness of the polar character. However, the simple picture of equating the sign change in the CD-ARPES spectra with a polar OAM texture is challenged when inspecting the photon-energy dependence (Fig.~\ref{fig:cd_photon}). Taking PdGa(111) as test case, there is a profound evolution of the CD-ARPES dispersions along the X-R-X direction when increasing the photon energy from $\hbar\omega=360$~eV (Fig.~\ref{fig:cd_photon}\textbf{a}) towards higher soft-X-ray energies (Fig.~\ref{fig:cd_photon}\textbf{d}). Strikingly, besides an overall sign inversion of the circular dichroism when going from $\hbar \omega$=360 eV to 547 eV, at even higher photon energies of $\hbar \omega$=767 eV and $\hbar\omega$=1021 eV the sign change of the CD-ARPES signal between positive and negative momenta with respect to the R point along the $k_x$ direction seems to vanish completely. This unexpected photon energy dependence shows that CD-ARPES is not a direct probe of the total OAM associated with the Bloch wavefunction, and raises the question if there is any deeper connection between the predicted OAM monopole and the polar CD-ARPES spectrum observed at lower photon energies. We will see in the following that this is indeed the case, and that the observed loss of sign change in the CD-ARPES signal along the $k_x$ direction at higher photon energies is caused by photon-energy dependent interference of local polar OAM textures from different atoms in the unit cell during the photoemission process.

Our simulations (Fig.~\ref{fig:cd_photon}\textbf{e}--\textbf{h}) capture the essence of the evolution of the CD-ARPES spectra with the photon energy (albeit the deviations are larger at high photon energies). In particular, the overall sign reversal upon increasing the photon energy from $\hbar\omega=360$~eV to 565 eV (Fig.~\ref{fig:cd_photon}\textbf{e},\textbf{f}) is clearly visible, while the loss of the sign-change along the $k_x$ axis at $\hbar\omega=1021$~eV is also reproduced. Upon closer inspection of the simulated CD-ARPES Fermi-surfaces in (Fig.~\ref{fig:cd_photon}\textbf{i}--\textbf{l}), the effect of increasing the photon energy manifests as \emph{rotation} of the polar circular dichroism texture over the Fermi-surface, i.e. the CD-ARPES still changes sign for positive and negative momenta with respect to R, but not symmetrically around the $k_x=0$ plane. At higher photon energy, the $k_y=0$ plane does not cut through the polar texture, resulting in the apparent loss of the characteristic sign change of the circular dichroism for positive and negative momenta in the experimental spectra (Fig.~\ref{fig:cd_photon}\textbf{c}--\textbf{d}).

\subsection*{Microscopic analysis of orbital texture and circular dichroism}
While the direct proportionality between the predicted OAM texture projected on the incident light direction and the CD-ARPES spectra is no longer valid at higher photon energies, the simulated CD-ARPES Fermi surfaces universally retain a polar structure as the key feature, which can be linked to the predicted polar OAM texture around the multifold fermion at R. To understand this link, we analyze our simulations in detail. Starting from Fermi's Golden rule, the photoemission intensity with left-handed (+) and right-handed (-) circularly polarized light can be described as
\begin{align}
    \label{eq:fermi_golden}
    I^{(\pm)}(\vec{k}_\parallel, E) = \sum_\alpha \sum_{k_\perp} |M^{(\pm)}_\alpha(\vec{k}_\parallel,E)|^2 g(\varepsilon_\alpha(\vec{k}) + \hbar\omega - E) \ ,
\end{align}
where $M^{(\pm)}_\alpha(\vec{k}_{\parallel},E)$ denotes the matrix element, $g(\omega)$ denotes a broadened Dirac delta function containing the band energies ($\varepsilon_\alpha(\vec{k})$), the photon energy $\hbar\omega$, the in-plane final state momentum $\vec{k}_\parallel$,  and the final-state energy $E$. The out-of-plane momentum $k_\perp$ in Eq.~\eqref{eq:fermi_golden} is determined from $E$ and $\vec{k}_\parallel$ through
\begin{align}
    \label{eq:koutofplane}
    \frac{\vec{k}_\parallel^2}{2} + \frac{k_\perp^2}{2} = E - \Phi \ ,
\end{align}
where $\Phi$ denotes the work function.

We can express the key ingredient for computing CD-ARPES -- the photoemission matrix element with respect to left-hand / right-hand circular polarization photons ($M^{(+)}_\alpha(\vec{k},E)$ / $M^{(-)}_\alpha(\vec{k},E)$) -- in terms of the underlying orbitals. Constructing an effective Wannier Hamiltonian from first-principle density-functional theory (DFT) calculations provides us with the optimal representation of Bloch wave-functions in terms of these orbitals (see Methods). For PdGa, the Wannier model includes $d$ orbitals localized on Pd sites and $p$ orbitals at the Ga sites. Ignoring effects related to the finite escape depth~\cite{moser_experimentalists_2017} (which mostly lead to additional broadening) of the photoelectrons, the circular dichroism can be expressed in terms of these orbitals in the bulk unit cell as
\begin{align}
\label{eq:cd_tensor}
\mathrm{CD}(\vec{k}_\parallel,E) &\propto |M^{(+)}_\alpha(\vec{k},E)|^2 - |M^{(-)}_\alpha(\vec{k},E)|^2 \nonumber \\ &
\approx \sum_{j j^\prime} C^*_{j\alpha}(\vec{k}) T_{j j^\prime}(\vec{k}) C_{j^\prime\alpha}(\vec{k}) \ .
\end{align}
Here, $C_{j\alpha}(\vec{k})$ describes the projection of Bloch state $|\psi_{\vec{k}\alpha}\rangle$ onto the orbital $j$, while the tensor $T_{j j^\prime}(\vec{k})$ incorporates the details photoemission matrix elements with respect to all orbitals. The expansion in terms of the orbitals~\eqref{eq:cd_tensor} allows us to dissect the various contributions to the circular dichroism. In particular, we can distinguish orbital-resolved, intra-atomic and interference contributions.

\begin{figure*}[t]
    \includegraphics[width=1.0\textwidth]{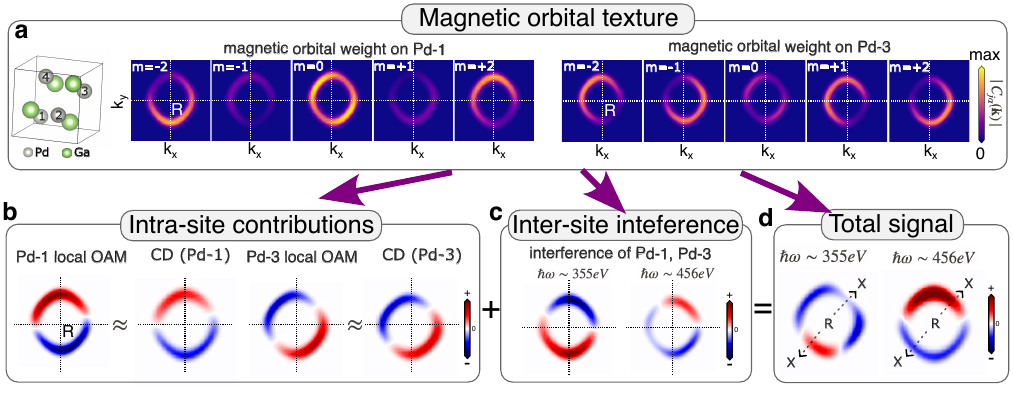}
    \caption{\textbf{Orbital-resolved analysis of the circular dichrosim.} \textbf{a} Texture of magnetic Pd-$d$ orbitals, at the two representative atomic sites Pd-1 and Pd-3, for $k_z=0$ close to the R point. The quantization axis is identical to the incident light. \textbf{b} Intra-site contributions from the Pd-$d$ orbitals, showing the qualitative equivalence of local OAM and the intra-atomic contributions to the circular dichroism. \textbf{c}
    Inter-site interference contribution, here exemplified for the interference channel between the Pd-1 and Pd-3 atom. \textbf{d} Total bulk signal, obtained by summing all intra-site and interference contributions.}
    \label{fig:illustration}
\end{figure*}

Focusing on the region close to the multifold fermion (we choose $0.2$~eV above the node), we find that mostly the Pd-$d$ orbitals contribute to the ARPES intensity. Each of the four Pd atoms carries momentum-dependent local OAM dictated by the chiral orbital texture of the magnetic quantum number $m=-2,\dots, 2$ with respect to the incidence direction (Fig.~\ref{fig:illustration}\textbf{a}). Let us first simulate the scenario where only a single selected Pd site contributes to the CD-ARPES signal, which can be simulated by replacing $j\rightarrow m$ in Eq.~\eqref{eq:cd_tensor} for the specific Pd atom and omitting all other terms, which we call local CD-ARPES. In this case, the \emph{local} OAM is directly reflected in the (local) circular dichroism (illustrated in Fig.~\ref{fig:illustration}\textbf{b}). If we now include all four Pd atoms but exclude inter-atomic interference effects, the total CD-ARPES signal is given by the sum of the dichroism originating from each of the Pd atoms, independent of the photon energy. This already complicates the interpretation, as the local OAM of each of Pd atoms exhibits a slightly different texture (exemplified in Fig.~\ref{fig:illustration}\textbf{b}) for two different Pd atoms).

Finally, including inter-orbital interference terms -- which have their specific polar texture -- gives rise to a complex photon energy-dependent pattern (see an example of two-atom interference in Fig.~\ref{fig:illustration}\textbf{c}). The dependence on the photon energy can be understood by geometric arguments: the interference terms include a factor $T_{j j^\prime}(\vec{k})\propto e^{-i\vec{k}\cdot(\vec{r}_{j} - \vec{r}_{j^\prime})}$ with $\vec{r}_{j}$ denoting the position of the Pd atoms in the unit cell.
As the Pd atoms are located at different vertical positions $z_{j} = [\vec{r}_{j}]_z$ in the unit cell and out-of-plane photoelectron momentum $k_\perp$ is photon-energy dependent (Eq.~\eqref{eq:koutofplane}) , the interference terms attain a dependence on the photon energy.

Combining all the contributions (Fig.~\ref{fig:illustration}\textbf{d}) seems to impede the direct extraction of the OAM from the measured circular dichroism. Indeed, there is no simple specific rule - such as projecting the OAM along the incoming light direction - that governs the orientation of the resulting CD-ARPES iso-energy surfaces. However, the key ingredient determining the OAM -- the asymmetric momentum-space texture of magnetic $d$ orbitals (Fig.~\ref{fig:illustration}(a)) -- manifests itself in \emph{all} contributions to the simulated circular dichroism. Because every single contribution exhibits a polar orbital texture, the resulting total CD-ARPES signal formed by interference will (except for an accidental cancellation at selected photon energies) generically show a polar texture as well. Hence, the polar structure of the circular dichroism that we observed for lower photon energies around the R point in Fig.~\ref{fig:cd_orientation} across different chiral crystals is a universal and relatively robust hallmark of the presence of OAM monopoles in our samples.

To further underpin the key role of the chirality of the crystal structure for the formation of OAM monopoles, we have performed CD-ARPES simulations for the hypothetical closest crystal structure of PdGa possessing inversion symmetry in the bulk. As a result, the orbitals no longer carry a magnetic moment, and the weight of the $\pm m$ orbitals becomes equal. The band structure is still qualitatively similar, enabling direct comparison. While the simulated circular dichroism does not vanish due to the broken inversion symmetry at the surface, the polar texture is completely lost (see supplemental material~\cite{supplement}).

\section*{Discussion}
While the OAM clearly manifests in in CD-ARPES experiments, extracting the precise texture of OAM -- even on a qualitative level -- is difficult. In particular, orbital and geometric interference effects that are always present in complex quantum materials break the direct correspondence of the circular dichroism and the OAM. For some materials, symmetry arguments can help identify characteristic sign changes in momentum space~\cite{unzelmann_momentum-space_2021,cho_studying_2021}, albeit only close to high-symmetry points. In general and especially for chiral materials as studied here, symmetries can not be exploited, and the OAM texture is obscured, making it more challenging to directly detect the presence of OAM monopoles from CD-ARPES spectra.

Our work overcomes these challenges by comprehensive measurements varying the crystal orientation and the photon energy, providing a detailed fingerprint of the underlying orbital texture. In particular, the monopole texture of the OAM around the R point is driven by a polar orbital texture, which manifests in a photon-energy-dependent CD-ARPES signal  however with a ubiquitous polar structure. Through the robustness of the polar circular dichroism, we have identified an unambiguous experimental signature of OAM monopoles in chiral materials.

This achievement raises the question if other OAM textures, such as sources and sinks encountered at nodal lines~\cite{bian_topological_2016,shao_dirac_2019} or anti-crossings of Bloch bands that appear in many other quantum materials, can also be identified from CD-ARPES measurements. Based on our current findings, we expect that in the vicinity of such special points, there should also exist a characteristic texture of magnetic orbitals that should lead to a characteristic local circular dichroism. As supported by our analysis of the CD-ARPES signal in terms of the interference channels, sign changes in the circular dichroism that are robust against varying photon energy or the crystal orientation are expected to provide a unique fingerprint of such OAM textures. Thus, our work paves the path to directly measuring key features of momentum-resolved OAM textures that could lead to interesting magneto transport and optical responses in a variety of quantum materials utilizing CD-ARPES.

\clearpage

\section*{Acknowledgment}
 We thank Gerhard H Fecher (MPI CPfS Dresden) for support with numerical calculations during the early stages of the project. We acknowledge Swiss Light source for beamtime on ADRESS  under Proposals No.~20212108, 20201889, and 20160611.
 J.A.K.~acknowledges support by the Swiss National Science Foundation (SNF-Grant
No.~P500PT\_203159 ).
M.S. and Y.Y. acknowledge support from SNSF Ambizione Grant No. PZ00P2 193527.  This research was supported by the NCCR MARVEL, a National Centre of Competence in Research, funded by the Swiss National Science Foundation (grant number 205602). K. M. acknowledges the Department of Atomic Energy (DAE), Government of India, for the funding support via Grant No. 58/20/03/2021- BRNS/37084; Department of Science \& Technology (DST), Govt of India, via Core Research Grant (CRG) with CRG/2022/001826 and the Max Planck Society for the funding support under the Max Planck–India partner group project.

\section*{Competing Interest}
The authors declare no competing interest.

\section*{Data availability}
The data of this study will be made available on the Open Research Data Repository of the Max Planck Society. (reserved DOI: \href{https://doi.org/10.17617/3.PILCPQ}{10.17617/3.PILCPQ} )

\section*{Methods}

\subsection*{Crystal growth}

PdGa and PtGa single crystals were grown from the melt with the self-flux technique as described in Refs.~\cite{schroter_observation_2020,yao_observation_2020}. After aligning the crystals at room temperature with a white-beam backscattering Laue X-ray setup, the surfaces along different high symmetry directions were polished.

\subsection*{ARPES measurements}
For the ARPES experiments, the crystal surfaces were cleaned in-situ with multiple sputtering (Ar$^+$, \SI{1}{keV}, \SI{20}{min}, \SI{1e-5}{mbar}) and annealing ($>\SI{870}{K}$, $\gtrsim\SI{20}{min}$), as described in Refs.~\cite{schroter_observation_2020,krieger_parallel_2022}.
Soft X-ray (SX-ARPES) experiments were performed at the SX-ARPES endstation~\cite{strocov_soft-x-ray_2014} of the ADRESS beamline~\cite{strocov_high-resolution_2010} at the Swiss Light Source, Switzerland, using a PHOIBOS-150 (SPECS) analyzer with the sample held around \SI{20}{K} and at a pressure better than \SI{2e-10}{mbar}.
The angular resolution was about \SI{0.1}{\degree}, and the combined analyzer and beamline energy resolution ranged from \SI{60}{meV} to \SI{180}{meV}
for photon energies between \SI{360}{eV} and \SI{1.021}{keV}.
Measurements were acquired with LCP and RCP polarization and  normalized by the total intensity ($I^\pm_{\rm{tot}}$) within the full measured range of the displayed 2D spectra, before calculating the dichrosim and total intensity:
\begin{align}
    CD=\frac{I^+}{I^+_{\rm{tot}}}-\frac{I^-}{I^-_{\rm{tot}}},\quad
    I_{\rm{tot}}=\frac{1}{2}\left(\frac{I^+}{I^+_{\rm{tot}}}+\frac{I^-}{I^-_{\rm{tot}}}\right).
\end{align}
In addition, for the spectral cuts (Fig.~\refsubfig{fig:cd_photon}{(a-d)}) $I_{\rm{tot}}$ was subsequently corrected for the analyzer transmission (estimated by energy integrating the spectra), and angle integrated background subtracted. The intensity in all Fermi surface iso-energy maps was integrated within $\pm\SI{100}{meV}$ of the Fermi level.

\subsection*{DFT calculations and Wannier tight-binding hamiltonian of PdGa/PtGa}
The ground-state band structure of PdGa/PtGa was obtained from DFT calculations with the QUANTUM ESPRESSO code~\cite{giannozzi_quantum_2009}. We choose PBE~\cite{perdew_generalized_1996} generalized gradient approximation (GGA) for the exchange-correlation functional without spin-orbit coupling.
We constructed Wannier functions for the bulk structure of both compounds with the Wannier90 code~\cite{pizzi_wannier90_2020}. We used the projective Wannier function approach without maximal localization to optimize the match of the Wannier functions with atom-like orbitals with well-defined angular quantum numbers. From the bulk Wannier Hamiltonian we constructed a slab supercell of 15 layers for the (001) and 20 layers for the (111) surface. The slab Hamiltonian is then used to perform the Wannier-ARPES calculations.

\subsection*{Wannier-ARPES simulations}
To evaluate the photoemission intensity~\eqref{eq:fermi_golden}, we need to compute the matrix elements $M_\alpha(\vec{k}_\parallel,E) = \langle \chi_{\vec{k_{\parallel}},E} | \vec{e}\cdot \hat{\vec{r}} | \psi_{\vec{k}\alpha}\rangle$ with respect to a given light polarization $\vec{e}$. Here we represent the light-matter coupling in the dipole gauge. The dipole operator $\hat{\vec{r}}$ is evaluated using the modern theory of polarization~\cite{resta_quantum-mechanical_1998}, avoiding any ill-definedness (see ref.~\cite{schuler_polarization-modulated_2022}). We further apply the atomic-centered approximation (ACA), which reduces the calculation of the photoemission matrix element to a summation over atomic transition amplitudes:
\begin{align}
    \label{eq:matel_full}
    M_\alpha(\vec{k}_\parallel,E) = \sum_{jl} C_{jl\alpha}(\vec{k_{\parallel}})e^{-i\vec{k}\cdot\vec{r}_{jl}} e^{z_{jl}/\lambda} M^{\mathrm{(ACA)}}_{j}(\mathbf{k}_\parallel, E) \ .
\end{align}
Here, $C_{jl\alpha}(\vec{k_{\parallel}})$ are the coefficients of Wannier orbital $j$ in slab layer $l$, while $\vec{r}_{jl}$ denotes the position of the Wannier center in the system (the $z$-projection is given by $z_{jl}$). The mean-free path of the photoelectrons in the crystal $\lambda$ determines the $k_\perp$-broadening in the simulations. For clarity we focus on the case $\lambda\rightarrow \infty$.

The atomic transition amplitudes are defined as
\begin{align}
    \label{eq:matel_aca}
    M^{\mathrm{(ACA)}}_{j}(\mathbf{k}_\parallel, E) = \int d\vec{r}\, \chi^*_{\vec{k_{\parallel}},E}(\vec{r}) \vec{e}\cdot\vec{r} \phi_j(\vec{r}) \ ,
\end{align}
where $\phi_j(\vec{r})$ is the Wannier function of orbital $j$. Approximating the orbitals as atom-like wave-functions, $\phi_j(\vec{r}) = R_j(r) Y_{\ell_j m_j}(\Omega_{\vec{r}})$ and employing the plane-wave approximation to the final states, $\chi_{\vec{k_{\parallel}},E}(\vec{r}) \approx e^{i\vec{k}\cdot \vec{r}}$ then allows us to compute the atomic matrix elements~\eqref{eq:matel_aca} efficiently in terms of Clebsch-Gordan coefficiens. The remaining unknown ingredients are radial integrals
\begin{align}
    \label{eq:radial_int}
    I^{\pm}_{j}(E) = \int^\infty_0 dr\, r^3 j_{\ell_j \pm 1}(k r) R_j(r) \ .
\end{align}
The energy dependence of the radial integrals is available from independent calculations using the the Korringa-Rostoker-Kohn (KKR) formalism~\cite{ebert_calculating_2011}. In practice, we parameterize the radial functions $R_j(r)$ as Slater-type wave-functions such that the energy dependence of Eq.~\eqref{eq:radial_int} matches the the KKR calculations for the relevant photon energies.
All calculations are performed using the \textsc{dynamics-w90} code~\cite{michaelschueler_dynamics-w90_2022}.

In Fig. \ref{fig:illustration}\textbf{b}--\textbf{d}, the CD is calculated with the simplified photoemission model Eq.\eqref{eq:cd_tensor} using the bulk Wannier Hamiltonian. The simulation is performed at $k_{\perp}=\pi/a$ with a map on different $\mathbf{k}_{\parallel}$, where $a$ is the cubic lattice constant.  The result qualitatively reproduces the measured circular dichroism. In particular, the cut along the X-R-X direction in Fig.~\ref{fig:illustration} \textbf{d} is in excellent agreement with the CD-ARPES map in Fig.~\ref{fig:cd_orientation}\textbf{j}. To improve the quantitative agreement and obtain the results presented Fig.~\ref{fig:cd_orientation} and \ref{fig:cd_photon}, the presence of the surface needs to be included, which is achieved by simulating ARPES for an entire slab. While the details of the spectra change, the salient polar texture is only weakly affected, underlining the orbital picture outlined above.

\onecolumngrid
\clearpage
\begin{center}
\textbf{\protect\large{
\textit{Supplemental Information:}
Controllable orbital angular momentum monopoles in chiral topological semimetals}
}
\end{center}
\setcounter{equation}{0}
\newcounter{FiguresInMainText}
\setcounter{FiguresInMainText}{\value{figure}}
\setcounter{table}{0}
\makeatletter
\renewcommand{\theequation}{S\arabic{equation}}
\renewcommand{\thefigure}{S\the\numexpr\value{figure}-\value{FiguresInMainText}}
\renewcommand{\bibnumfmt}[1]{[S#1]}
\renewcommand{\citenumfont}[1]{S#1}

\tableofcontents
\let\addcontentsline\oldaddcontentsline

\section{Out-of-plane dispersion}
\begin{figure*}[hbt]
    \includegraphics[width=0.7\linewidth]{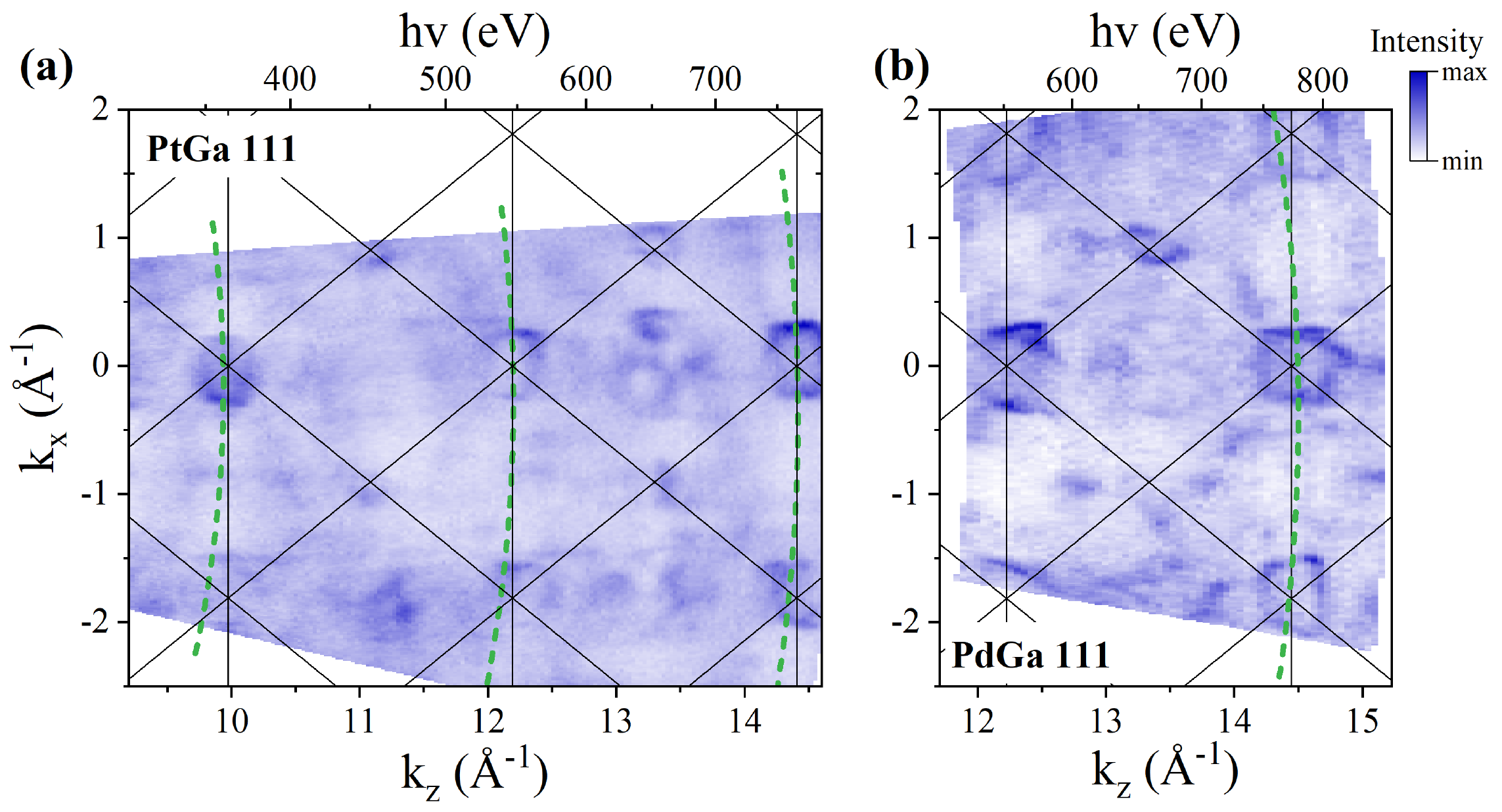}
  \caption{\textbf{Out-of-plane dispersion on the 111 surfaces.} (a) Fermi surface map with the top axis showing the photon energy at $k_x=\SI{0}{\angstrom^{-1}}$. (b) The same for PdGa 111.
  The expected Brillouin-zone boundaries are drawn as black lines. The green dashed lines show the constant photon energy cuts corresponding to the Fermi surfaces and spectra shown in the main paper.
  }
\label{fig:kz}
\end{figure*}

The out-of-plane momentum corresponding to a photoelectron with kinetic energy $E_{\mathrm{kin}}$ was calculated as:
\begin{equation}
{k}_{z}=\sqrt{\frac{2m_{\mathrm{e}}\left(E_{\mathrm{kin}}-V_{000}
\right) } { \hbar^2 }-{k}_{\parallel} }+{p}_{\gamma,z},
\end{equation}
where $m_{\mathrm{e}}$ is the electron mass, ${k}_{\parallel}$ the
in-plane momentum, $V_{000}$ the inner potential and
${p}_{\gamma,z}$ the component of the photon momentum ($h\nu/c$)
perpendicular to the sample surface. In particular, we used the expressions specifically derived for the specific measurement geometry of ADRESS in   Ref.~\cite{strocov_soft-x-ray_2014SI}. We have estimated
the inner potential of PtGa and PdGa to be
$V_{000}\approx\SI{12}{eV}$ by comparing the measured band structure with the expected periodicity of
the Brillouin zone shown as solid lines in Fig.~\ref{fig:kz}.

\section{P\lowercase{d}G\lowercase{a}(001) photon energy dependent circular dichroism}
\begin{figure*}
    \includegraphics[width=0.6\textwidth]{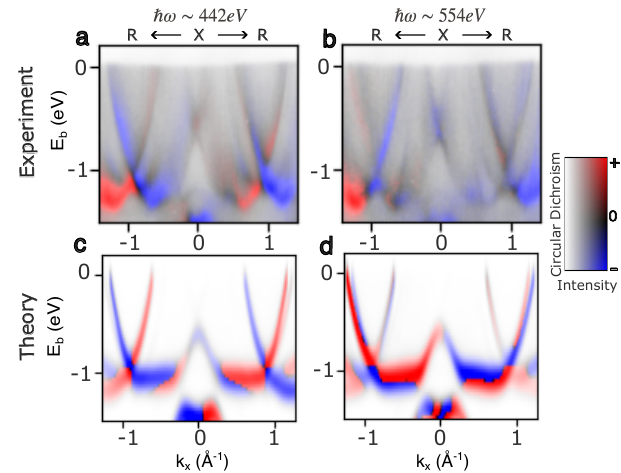}
    \caption{\textbf{CD photon energy dependence in PdGa(001)} \textbf{a},\textbf{b} Experimental and \textbf{c},\textbf{d} simulated CD in PdGa(001) agree well with each other.}
    \label{fig:pdga_wdep}
\end{figure*}

In Fig.~\ref{fig:pdga_wdep}, the photon energy dependence of CD in PdGa(001) also shows agreement between experiment and theory, just like in PtGa(111) (Fig.~3 in the main text). The data shows the CD sign change as a function of photon energy and can be understood as the "rotation" of polar CD signal (Fig.~4 in the main text).

\section{Symmetry relation of circular dichroism in two enantiomers}

Suppose $H_A$ and $H_B$ are Hamiltonian for enantiomers A and B. We set the mirror plane as the same as the scattering plane, which is the y-z plane in our coordinate system. As a result, mirror symmetry operator $\hat{M_x}$ relates the two enantiomers. Therefore we have
     \begin{align}
         &\hat{M_x}^{\dagger}H_A(\mathbf{k})\hat{M_x} = H_B(\hat{M_x}\mathbf{k})\\
         &\hat{M_x} \ket{\psi_{\alpha}^{A}(\mathbf{k})} = \ket{\psi_{\beta}^{B}(\hat{M_x}\mathbf{k})}
     \end{align}

     , where $\alpha$ and $\beta$ label Bloch bands for A/B.

     On the other hand, the dipole matrix elements for A/B can be written as
     \begin{align}
         &\langle \chi^{B} | \hat{\mathbf{r}} | \psi_{\beta}^{B}(\hat{M_x} \mathbf{k}) \rangle \\
         =& \langle \chi^{A} | \hat{M_x^{\dagger}}\hat{\mathbf{r}}\hat{M_x} | \psi_{\alpha}^{A}(\mathbf{k}) \rangle \\
         =& \langle \chi^{A} | (-\hat{\mathbf{x}}, \hat{\mathbf{y}}, \hat{\mathbf{z}}) | \psi_{\alpha}^{A}(\mathbf{k}) \rangle
     \end{align}
     , where $\ket{\chi^{A(B)}}$ denotes the photoelectron final state for enantiomer A(B) with the relation $\ket{\chi^{B}} = \hat{M_x} \ket{\chi^{A}}$. Now with our exerimental geometry (light is 20$^o$ away from crystal plane), the light-matter interacting matrix elements in dipole gauge are simply
     \begin{align}
         D_B^{+}(\hat{M_x}\mathbf{k}) &= \mathbf{e^{+}} \cdot \langle \chi^{B} | \hat{\mathbf{r}} | \psi_{\beta}^{B}(\hat{M_x}\mathbf{k}) \rangle\\
         &= +i D^{Bx}(\hat{M_x} \mathbf{k}) + sin(20^{o}) D^{By}(\hat{M_x} \mathbf{k}) + cos(20^{o}) D^{Bz}(\hat{M_x} \mathbf{k}) \\
         &= -i D^{Ax}(\mathbf{k}) + sin(20^{o}) D^{Ay}(\mathbf{k}) + cos(20^{o}) D^{Az}(\mathbf{k}) = D_A^{-}(\mathbf{k})
     \end{align}
     \begin{align}
         D_B^{-}(\hat{M_x}\mathbf{k}) &= \mathbf{e^{-}} \cdot \langle \chi^{B} | \hat{\mathbf{r}} | \psi_{\beta}^{B}(\hat{M_x}\mathbf{k}) \rangle\\
         &= -i D^{Bx}(\hat{M_x} \mathbf{k}) + sin(20^{o}) D^{By}(\hat{M_x} \mathbf{k}) + cos(20^{o}) D^{Bz}(\hat{M_x} \mathbf{k}) \\
         &= +i D^{Ax}(\mathbf{k}) + sin(20^{o}) D^{Ay}(\mathbf{k}) + cos(20^{o}) D^{Az}(\mathbf{k}) = D_A^{+}(\mathbf{k})
     \end{align}
     , where $\mathbf{e}^{+(-)}$ represents right(left) circularly polarization. $D^{A(B)\mu} = \langle \chi^{A(B)} | \hat{\mu}| \psi_{\alpha(\beta)}^{A(B)}(\mathbf{k}) \rangle $ is the matrix element for cartesian direction $\mu$. Finally we find that
     \begin{align}
        \label{eq:sym_cd}
         CD_A(\mathbf{k}) = -CD_B(M_x \mathbf{k}).
     \end{align}

     In Fig.~2\textbf{a} of the main text, the dichroic Fermi surface of the enantiomer A is measured with light direction $60^o$ differed from enantiomer B (Fig.~2\textbf{b}).
     The matching simulation in Fig.~2\textbf{g} and \textbf{h} uses the same enantiomer with corresponding light geometry. Next we transform CD in enantiomer A (Fig.~2\textbf{g}) using the symmetry relation (\ref{eq:sym_cd}). The agreement between experiment and simulation confirms the symmetry analysis.

\section{Circular dichroism in inversion symmetric P\lowercase{d}G\lowercase{a}}
\begin{figure*}
    \includegraphics[width=1.0\textwidth]{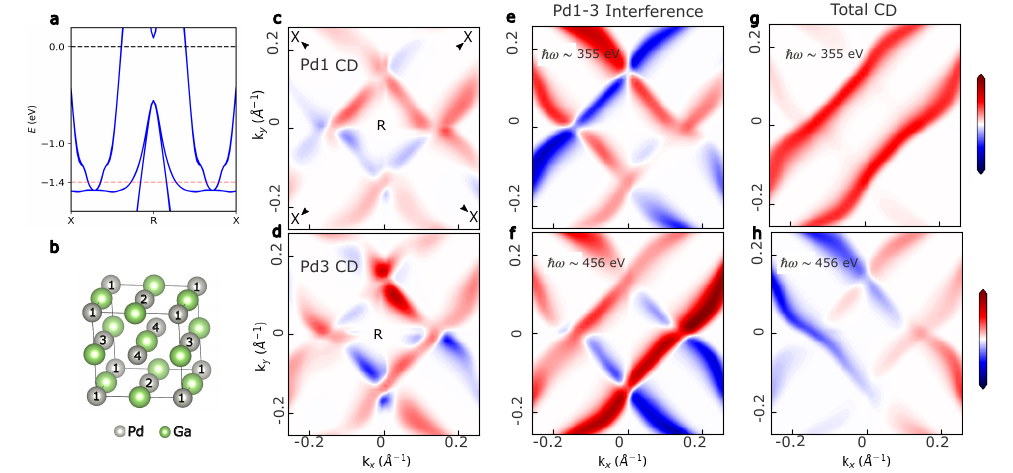}
    \caption{\textbf{Inversion-symmetric PdGa simulation} \textbf{a} Band structure of \textbf{b} the inversion-symmetric PdGa. \textbf{c},\textbf{d} Local CD from Pd1 and Pd3 atoms. \textbf{e},\textbf{f} inter-site interferece between Pd1 and Pd3 at different photon energies. \textbf{G},\textbf{H} Total CD signal with summation over every intra-site term and inter-site interference terms.}
    \label{fig:inversion}
\end{figure*}

The observed photon energy dependent polar CD is a direct manifestation of OAM monopoles and chiral orbital texture, as discussed in the main text. To support this, we manually restore inversion symmetry in PdGa as in Fig.~\ref{fig:inversion}\textbf{B} and simulated the corresponding CD. In the bulk band structure (Fig.~\ref{fig:inversion}\textbf{A}), the crossings at R points are no longer Weyl points or OAM monopoles. As a result, local CD (Fig.~\ref{fig:inversion}\textbf{C}-\textbf{D}), inter-site interference terms (Fig.~\ref{fig:inversion}\textbf{E}-\textbf{F}), and the total CD (Fig.~\ref{fig:inversion}\textbf{G}-\textbf{H}) lose the polar nature.

\section{Global and local Orbital Angular Momentum}

Orbital angular momentum (OAM) for periodic solids is rigorously defined in terms of the modern theory of magnetization~\cite{xiao_berry_2010SI}:
   \begin{equation}
        \label{eq:modern_oam2}
        L_{\alpha}^{\mu}(\mathbf{k}) = -i \epsilon_{\mu \nu \gamma} \sum_{\alpha \neq \alpha'} (\varepsilon_{\alpha'}(\mathbf{k}) - \varepsilon_{\alpha}(\mathbf{k})) A_{\alpha \alpha'}^{\nu}(\mathbf{k}) A_{\alpha' \alpha}^{\gamma}(\mathbf{k}) \ .
    \end{equation}
Here, $\alpha$ denotes the band index, $\mu$ stands for the cartesian directions, and $\epsilon_{\mu\nu\gamma}$ is the Levi-Civita tensor. Besides the band structure $\varepsilon_\alpha(\mathbf{k})$, the Berry connections $A^{\nu}_{\alpha\alpha^\prime}(\mathbf{k}) = i \langle u_{\mathbf{k}\alpha}| \partial_\nu u_{\mathbf{k}\alpha^\prime}\rangle$ are the central ingredients to Eq.~\eqref{eq:modern oam2}. The modern theory of OAM is closely tied to Berry curvature~\cite{xiao_berry_2010SI}; indeed, if there are only two two bands, both quantities are proportional to each other~\cite{ma_chiral_2015}.

While the modern theory definition~\eqref{eq:modern oam2} defines a \emph{global} OAM that is proportional to the magnet moment of the crystal, ARPES experiments in the measurement geometry used in this work are sensitive to the \emph{local} OAM at the different atomic sites. In terms of the Wannier Hamiltonian, the local OAM is defined as
\begin{equation}
    \label{eq:modern oam2}
    L_{a,\alpha}^{\mu}(\mathbf{k}) = \sum_{m',m}  C_{(am)\alpha}^{*}(\mathbf{k})  L_{mm'}^{\mu} C_{(am')\alpha}(\mathbf{k}) \ ,
\end{equation}
where $m,m'$ run through magnetic orbitals localized at atomic site $a$. $L_{mm'}^{\mu}$ is the atomic OAM matrix elements, which depends on our choice of the quantization axis. Due to the broken translational invariance in the out-of-plane direction, the momentum derivative $\partial_z = \partial / \partial k_z$ reduces to the standard dipole operator in $z$-direction. As a consequence, the site-resolved circular dichroism corresponds to the local OAM at respective atomic site.

\begin{figure*}
    \includegraphics[width=1.0\textwidth]{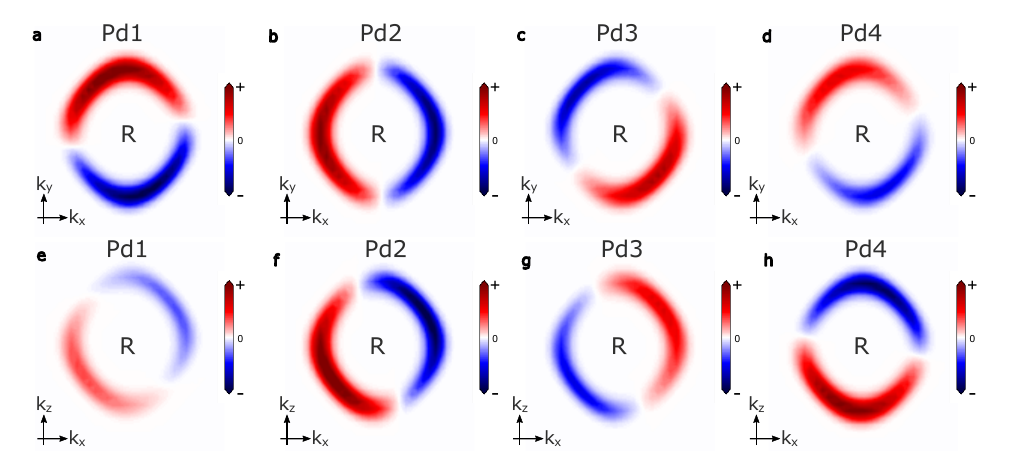}
    \caption{\textbf{Local OAM projected onto incoming light direction for Pd1-Pd4} Local OAM projection for \textbf{a} -- \textbf{d} $k_x - k_y$ plane and \textbf{e} -- \textbf{h} $k_x - k_z$ plane are calculatd at 0.2 eV above the node. \textbf{a} and \textbf{c} are identical to the local OAM of Pd1 and Pd3 in Fig. 4\textbf{b} of the main text.}
    \label{fig:local_pole}
\end{figure*}

In Fig.~1\textbf{b},\textbf{c} in the main text we show the global OAM from the modern theory~\eqref{eq:modern_oam2}. The global OAM -- and the Berry curvature -- exhibit the hedgehog texture in the vicinity of the R-points. This hedgehog structure is also imprinted onto the local OAM at each atomic site in the unit cell, which is reflected in the polar structure of the local OAM in  Fig.\ref{fig:local_pole}.

\section{Correspondence between local OAM and local CD}

    Fig.2 and Fig.3 in the main text demonstrate the remarkable agreement between the experiment and theory, which indicates the applicability of atomic centered approximation (ACA) in this work.
    In this section, we analyze how local intra-site CD computed with the ACA corresponds to the the local OAM, in the case of $d$ orbital character bands. We rewrite the expression of circular dichroism in the case of infinite escape depth $\lambda \rightarrow \infty$ as

    \begin{align}
        \label{eq:cd_simplified}
        \mathrm{CD}(\mathbf{k},E) &\propto \left|\sum_j C_{j\alpha}(\mathbf{k}) e^{-i\mathbf{k}\cdot\mathbf{r}_{j}} M^{\mathrm{(+)}}_{j}(\mathbf{k}, E) \right|^2 - \left|\sum_j C_{j\alpha}(\mathbf{k}) e^{-i\mathbf{k}\cdot\mathbf{r}_{j}} M^{\mathrm{(-)}}_{j}(\mathbf{k}, E) \right|^2 \nonumber \\
        &= \sum_{j j^\prime} C^*_{j\alpha}(\mathbf{k}) T_{j j^\prime}(\mathbf{k},E) C_{j^\prime\alpha}(\mathbf{k}) \ ,
    \end{align}

where the matrix element $M^{+}(\mathbf{k},E)$ [$M^{-}(\mathbf{k},E)$] denotes the matrix element in the ACA with respect to the right-hand circularly (+) or left-and circularly (-) polarized light. The index $j=(a,m)$ includes atom site index ($a$) and atomic orbital index ($m$). The matrix $T_{j j^\prime}(\mathbf{k},E)$ includes all the matrix element and experimental geometry effects, with the expression

    \begin{align}
        T_{j j^\prime}(\mathbf{k},E) = e^{-i\mathbf{k}\cdot(\mathbf{r}_j-\mathbf{r}_{j^\prime})} [M^{\mathrm{(+)}*}_{j}(\mathbf{k}, E)M^{\mathrm{(+)}}_{j'}(\mathbf{k}, E) - M^{\mathrm{(-)}*}_{j}(\mathbf{k}, E)M^{\mathrm{(-)}}_{j'}(\mathbf{k}, E)]
    \end{align}

    The full CD is then consist of intra-site local CD with the same atomic site $T_{amam'}$, and the inter-atomic interference terms with different atomic sites $T_{ama'm'}$ [CD simulation in Fig.4\textbf{b}--\textbf{c} in the main text].

      The relation between local OAM and local intra-site CD depends on the details of the tensor $T_{jj'}(\mathbf{k},E)$ and the atomic OAM matrix elements $L^{\mu}_{mm'}$. This also indicates that the correspondence between local OAM and CD is sensitive to the experimental geometry and the types of contributing orbitals. As an illustration, we calculate $T_{jj'}(\mathbf{k},E)$ with photon energy $\omega=400$ eV for d orbitals of a Pd atom, using the same experimental geometry as our PdGa(001) measurement. We can then relabel the CD tensor $T_{jj'}$ as $T_{mm'}$ with magnetic quantum numbers $m,m'=-2, \dots, +2$. At $\Gamma$ point, we find

     \begin{align}
        \label{eq:cd_tensor_d_gamma}
        T_{mm'} =
        \begin{bmatrix}
    -5.5 & 0.0 & 3.0i & -1.13+1.13i & 0.0\\
    0.0 & 0.46 & -0.66-0.66i & 0.0 & -1.13+1.13i\\
    -3.0i & -0.66+0.66i & 0 & -0.66-0.66i & -3.0i\\
    -1.13-1.13i  & 0 & -0.66+0.66i & -0.46 & 0.0 \\
    0 & -1.13-1.13i  & 3.0i & 0.0 & 5.5
    \end{bmatrix} \times 10^{-3} + O(10^{-5})
    \end{align}

     If the quantization axis of the OAM operator is set to be along incoming light direction, then we have
    \begin{equation}
        L^{exp}_{mm'} =
        \begin{bmatrix}
        -2 & 0 & 0 & 0 & 0\\
        0 & -1 & 0 & 0 & 0\\
        0 & 0 & 0 & 0 & 0\\
        0 & 0 & 0 & 1 & 0\\
        0 & 0 & 0 & 0 & 2
        \end{bmatrix}
    \end{equation}

    The proportionality between OAM and CD is then not guaranteed due to two reasons. First, $L^{exp}_{mm'}$ is diagonal, while $T_{mm'}$ has some non-zero off-diagonal terms. Second, $T_{m=m'=-2}$($T_{m=m'=+2}$) has opposite sign to $T_{m=m'=-1}$($T_{m=m'=+1}$). Fortuanely, we find that at large photon energies (few hundred eV), the final state momentum is dominated by the component perpendicular to the surface, which leads to the dominating absolute value of the matrix elements $|T_{m=m'=\pm 2}|$ compared to the other terms in $T_{mm'}$. This means that if the local OAM is mostly determined by $m=\pm 2$ orbitals, the intra-site CD measurement is mostly proportional to the local OAM.

    This is indeed the case for PdGa. In Fig.4\textbf{a} of the main text, we show the local OAM of both Pd1 and Pd3 are roughly proportional to the difference between weight of m$=\pm 2$ orbitals. As a result, the intra-site CD, which is calculated with the simplified phtoemission model, is roughly proportional to the local OAM in Fig.4\textbf{b}.

\end{document}